# Textural properties of synthetic nano-calcite produced by hydrothermal carbonation of calcium hydroxide

G. Montes-Hernandez [*,a], A. Fernández-Martínez [a,b], L. Charlet [a], D. Tisserand [a], F. Renard [c,d]

[a] LGIT, University of Grenoble and CNRS, BP 53 X, 38420 Grenoble Cedex 9

[b] Institut Laue-Langevin, B.P. 156, 38042 Grenoble Cedex 9

[c] Physics of Geological Processes, University of Oslo, Norway

[d] LGCA, University of Grenoble and CNRS, BP 53 X, 38420 Grenoble Cedex 9

[*] Corresponding author: German Montes-Hernandez

E-mail address: German.MONTES-HERNANDEZ@obs.ujf-grenoble.fr

german_montes@hotmail.com

**Abstract**

The hydrothermal carbonation of calcium hydroxide ($Ca(OH)_2$) at high pressure of $CO_2$ (initial $P_{CO2}$ = 55 bar) and moderate to high temperature (30 and 90°C) was used to synthesize fine particles of calcite. This method allows a high carbonation efficiency (about 95% of $Ca(OH)_2$-$CaCO_3$ conversion), a significant production rate (48 kg/m$^3$h) and high purity of product (about 96%). However, the various initial physicochemical conditions have a strong influence on the crystal size and surface area of the synthesized calcite crystals. The present study is focused on the estimation of the textural properties of synthesized calcite (morphology, specific surface area, average particle size, particle size distribution and particle size evolution with reaction time,) using Rietveld refinements of X-ray diffraction spectra, BET measurements and, scanning and transmission electron microscope (SEM and TEM) observations. This study demonstrate that the pressure, the temperature and the dissolved quantity of $CO_2$ have significant effect on the average particle size, specific surface area, initial rate of precipitation, and on the morphology of calcium carbonate crystals. In contrast, these PTx conditions used herein have insignificant effect on the carbonation efficiency of $Ca(OH)_2$.
Finally, the results presented here demonstrate that nano-calcite crystals with high specific surface area ($S_{BET}$=6-10 m$^2$/g) can be produced, with a high potential for industrial applications such as adsorbents and/or filler in papermaking industry.

# 1. Introduction

Calcium carbonate is an inorganic compound that has been widely studied due to its abundance in nature as a mineral and biomineral. Calcium carbonate particles have three crystal morphologies and structures, which are generally classified as rhombic calcite, needle-like aragonite and spherical vaterite. Calcite belonging to Trigonal-Hexagonal-Scalenohedral Class is the most stable phase at room temperature under normal atmospheric conditions, while aragonite and vaterite belong to Orthorombic-Dipyramidal Class and Hexagonal-Dihexagonal Dipyramidal Class, respectively. The later are metastable polymorphs which readily transform into the stable calcite. The specific formation of one of the polymorphs of crystalline calcium carbonate particles depends mainly on the precipitation conditions, such as pH, temperature and supersaturation. Supersaturation is usually considered to be the main controlling factor [1]. Many experimental studies have been reported about the synthetic precipitation of the various forms of calcium carbonate and the conditions under which these may be produced, including the importance of initial supersaturation, temperature, pH and hydrodynamics. The effect of impurities and additives has also been well studied [see for ex. 2-15].

Limestone is the most common natural form of calcium carbonate. It has extensive industrial applications: it is one of the components of cements and several construction materials. However, the actual usefulness of calcium carbonate extends far beyond the current usages to which limestone is put. For instance, fine calcium carbonate particles are an active ingredient in commercially available antacid tablets. The calcite is also efficient filler in printing inks and papermaking industry [16].

The most commonly used industrial process for obtaining $CaCO_3$ powder involves the following steps: (a) the production of quicklime (CaO) and carbon dioxide by calcinations of limestone; (b) the transformation of quicklime to slaked lime slurry (a suspension of $Ca(OH)_2$ particles) by controlled addition of water; and finally (c) the carbonation reaction (Eq. 1), in which the $CO_2$ is bubbled through an aqueous slurry of slaked lime [17].

$$Ca(OH)_{2(s)} + CO_{2(aq)} \rightarrow CaCO_{3(s)} + H_2O \tag{1}$$

The carbonation reaction is the crucial step determining the textural properties (such as average particle size, particle size distribution, morphology and specific surface area) of the obtained product. For example, the morphology of the precipitated calcite, at typical temperatures of industrial process (between 30 and 70 °C), is normally the scalenohedral one bounded by the $\{21\bar{1}\}$ form. Synthetic scalenohedral calcite is generally produced through a batch carbonation method. The rhombohedral morphology, bounded by the {104} form, is usually precipitated by using solution routes, but rarely by the mentioned industrial process. The carbonation of $Ca(OH)_2$ suspension without the addition of additives allows to control the textural properties of calcite precipitates [17, 18-19]. Therefore, the development of new industrial carbonation routes for the production of calcite with different textural properties in the absence of expensive additives is of great interest. The industrial applications of $CaCO_3$ are mainly determined by the textural properties, such as average particle size, particle size distribution, morphology, specific surface area and/or polymorphism [17].

Recently, the hydrothermal carbonation of $Ca(OH)_2$ suspension in presence of compressed or supercritical $CO_2$ was proposed as a novel method to produce powdered calcite with potential interest to industrial applications [17, 20]. Then, in the current investigation, the carbonation of calcium hydroxide suspension at high pressure of $CO_2$ (initial $P_{CO2} = 55$ bar) and moderate to high temperature (30 and 90°C) was used to synthesize fine particles of calcite. In a previous study it was shown that the proposed method allowed a high carbonation efficiency (about 95% of $Ca(OH)_2$-$CaCO_3$ conversion), a significant production rate (48 kg/m$^3$h) and high purity of product (about 96%) [21]. In this study, it was also observed that the carbonation efficiency was not significantly affected by Pressure and Temperature (P-T) conditions after 24 h of reaction. This confirmed the low reactivity of molecular $CO_2$ on calcite dissolution [22] and precipitation [21]. In contrast, the initial rate of calcite precipitation was proportional to dissolved $CO_2$. For this case, it increased from 4.3 mol/h in the "90 bar – 90 °C" system to 15.9 mol/h in the "55 bar – 30 °C" system. This kinetic behaviour should have a direct influence on the texture properties, particularly on the average particle size, particle size distribution, specific surface area and crystal morphology. For this reason, the aim in the present study is mainly focused on the quantification of the textural properties of synthesized calcite by using Rietveld refinement of X-ray diffraction (XRD) spectra, Brunauer-Emmet-Teller (BET) specific surface area measurements and Scanning and Transmission Electron Microscope (SEM and TEM) observations. By varying the initial P, T, and molar fraction for calcite synthesis, different grain sizes could be produced. For this study, nanoparticles are defined as those characterized by 2 to 3 dimensions lower than 300 nm.

**2. Materials and methods**

*2.1 Synthesis of calcite (stirred reactor)*

One litre of high-purity water with electrical resistivity of 18.2 MΩ·cm and 74.1 g of commercial calcium hydroxide (provided by Sigma-Aldrich) with 96% chemical purity (3% $CaCO_3$ and 1% other impurities) were placed in a titanium reactor (autoclave with internal volume of two litres). The hydroxide particles were immediately dispersed with mechanical agitation (400 rpm). The dispersion was then heated to 90°C with a heating system adapted to the reactor. When the dispersion temperature was reached, 80.18 g of $CO_2$ (provided by Linde Gas S.A.) were injected in the reactor and the total pressure in the system was immediately adjusted to 90 bar by argon injection (see Figure 1a). Under these P-T conditions, the vapour phase consists mainly of an Ar + $CO_2$ mixture with the $CO_2$ in a supercritical state (Figure 1b). In order to evaluate the precipitation (or production) rate, five different reaction times were considered (0.25, 0.5, 4, 15 and 24 h). The experiments were also carried out at 30°C and 55 bar for reaction durations of 0.25, 4 and 24 h. For this second case, 96.05 g of $CO_2$ were initially injected in the reactor. At 55 bar and 30°C, the vapour phase consists mainly of gaseous $CO_2$ (Figure 1b). A complete description on the experiments was reported by Montes-Hernandez et al. [21].

Additionally, a semi-batch system (sampling with time) was performed in order to measure the pH (using MA235 pH/ion analyzer) and calcium concentration (using ICP Perkin Elmer Optima 3300 DV) in filtered solutions. For this case, about 25 ml of suspension were sampled in the reactor as a function of time (0, 2, 6, 10 and 30 minutes) during calcite precipitation. The P-T conditions were the same above cited, but, only 3g of calcium hydroxide were placed in the reactor. This allows better particle dispersion and a faster dissolution of calcium hydroxide. In addition, for these experiments, approximately 14.5 g of $CO_2$ were injected in the system. Note that the pH and Ca-concentration measurements were carried out at 25 °C after filtration, cooling and degasification of the solutions.

*2.2 Characterization of the solid particles size and specific surface area*

Morphological analyses of the solid products were performed by Scanning Electron Microscopy (SEM), with a HITACHI S-4800 microscope. Isolated fine particles (oriented on carbon Ni grids) of the starting material and products were also studied using a JEOL 3010 Transmission Electron Microscope (TEM) equipped with an energy dispersive X-ray analyser (EDS) to image the morphology of the particles and to identify the precipitated phases.

The specific surface area of powdered calcite (four samples) was estimated by applying the Brunauer-Emmet-Teller (BET) equation and by using 16.3 $Å^2$ for cross-sectional area of molecular $N_2$. The $N_2$ adsorption experiments were performed using a Micrometrics ASAP 2010 system.

*2.3 X-ray diffraction analysis of the solid phase*

X-Ray Powder Diffraction (XRD) analyses were performed using a Kristalloflex 810, SIEMENS diffractometer in Bragg-Brentano geometry. The XRD patterns were collected using Co k$\alpha_1$ ($\lambda_{k\alpha 1}$=1.7889 Å) and k$\alpha_2$ ($\lambda_{k\alpha 2}$=1.7928 Å) radiation in the range *2θ* = 5 - 80° with a step size of 0.02° and a counting time of 8 seconds per step.

The widths of the diffraction peaks in a diffraction pattern are mainly affected by two contributions: (1) the instrumental resolution and (2) microstructural effects related to the size of the crystallites or to the strains within their crystal structure. Both effects are convoluted in a diffraction pattern but can be separated by the use of convolution or deconvolution methods. A convolution method (Rietveld refinement) has been used in this study to account for size and strain effects. For this purpose, an X-ray diffraction pattern of a microstructure-free sample ($SiO_2$ in our case) has been measured, in order to isolate the instrumental resolution effect. In this way, further Rietveld refinements of the samples using size and strain models account only for microstructural effects.

In order to get a more detailed knowledge on the texture of the calcite particles, we have performed microstructural analyses within the Rietveld refinement of the X-ray diffraction data. The method is sensitive to the anisotropy of the crystallites, as a refinement of the full X-ray pattern is performed, which includes reflections of different reciprocal space directions. In addition, the combination of this crystallographic approach with the use of adsorption-based methods for the calculation of specific surfaces gives an idea of the crystallinity and aggregation state of the samples, as different entities are probed: the aggregation/agglomeration of different crystalline grains can decrease the porosity and thus prevent the access of the $N_2$ molecules to the entire surface, while the X-ray diffraction probes the coherent domains of the sample and yields and estimation of its crystallinity.

Rietveld refinement of the powder diffraction patterns has been carried out using the FullProf package (Windows version, February 2007) [23]. The pseudo-Voigt profile function of Thompson, Cox and Hastings [24] was used to fit the peak shapes. The parameters constituting the angular-dependent microabsorption correction were refined for each sample. Multiphase analysis was performed introducing a portlandite phase (calcium hydroxide) in the samples when the portlandite to calcite phase transition was not completely accomplished. Structural refinements of calcite were performed considering the R-3C space group and taking as initial values those obtained in the x-ray powder diffraction study reported by Maslen et al. [25]. Scale factor, zero point, cell dimensions, atomic coordinates and Debye Waller factors were refined. Background was refined by adjusting a fourth order polynomial. The instrumental contribution to peak broadening was determined with a $SiO_2$ sample [26]. The asymmetry parameters of the L. Finger correction S_L and S_D [27], and the Cagliotti parameters [28] were refined for the standard sample of $SiO_2$ and kept fixed during the refinements of the calcite patterns. Anisotropic size broadening was modeled in terms of spherical harmonics and the coherent domain average apparent size along each reciprocal lattice vector was calculated. Coherent domain average size was calculated averaging the resulting size of each of the reciprocal space distances measured. Volume and specific surface area of the coherent domains were calculated by assuming rectangular prisms with edges along the reciprocal space [100], [010] and [001] directions. The Cagliotti parameter U was refined in order to account for some isotropic strain in the Gaussian component of the peak profile. The strain is given in ‰: a strain of *x ‰* means a strain ratio of $\varepsilon = (d_i - d_f) / d_i \times 10000$, being $d_i$ and $d_f$ the strain-free crystallite size and the strained crystallite size respectively. Errors in size and strain have been estimated by performing refinements with different starting points, thus probing the stability of the result. Errors in size and strain are better than 10%.

## 3. Results and discussion

*3.1. Reaction mechanism of calcite precipitation*

The aqueous carbonation of Ca(OH)$_2$ described by the global reaction (1) is an exothermic process that concerns simultaneously the dissolution of Ca(OH)$_2$,

$$Ca(OH)_{2(s)} \xrightarrow{water} Ca^{2+} + 2OH^- \qquad (2)$$

and the dissociation of aqueous CO$_2$,

$$CO_{2(aq)} + H_2O \rightarrow CO_3^{2-} + 2H^+ \qquad (3)$$

these processes produce a fast supersaturation ($S_I$) of solution with respect to calcite,

$$S_I = \frac{(Ca^{2+})(CO_3^{2-})}{K_{sp}} > 1 \qquad (4)$$

where ($Ca^{2+}$) and ($CO_3^{2-}$) are the activities of calcium and carbonate ions in the solution, respectively, and K$_{sp}$ is the thermodynamic solubility product of calcite. Then, the nucleation stage (formation of nuclei or critical cluster) takes place in the system (see Fig. 2),

$$Ca^{2+} + CO_3^{2-} \rightarrow CaCO_3(nuclei) \qquad (5)$$

Finally, the crystal growth occurs spontaneously until the equilibrium calcite and the solution is reached (see for ex. Fig. 3),

$$CaCO_3(nuclei) \rightarrow CaCO_3(calcite) \qquad (6)$$

The metastable crystalline phases of CaCO$_3$, such as vaterite and aragonite, were not identified in the X-ray diffraction spectra during the Ca(OH)$_2$ carbonation process in our experiments. A small quantity of crystalline aragonite can be precipitated when the reactor is depressurized after the water-cooling stage (for more details, see [21]).

*3.2. Microscopic and BET measurements*

For this study, the presence of supercritical $CO_2$ did not have a clear effect on the $Ca(OH)_2$ carbonation process. In contrast, the P-T conditions had a significant effect on carbonation rate [21]. In fact, the precipitation rate is proportional to the quantity of dissolved $CO_2$. This justifies a higher rate of calcite precipitation at lower temperature. Consequently, we propose that the precipitation rate for these systems is linked to the particle size and calcite-crystal morphology. At higher precipitation rate (15.9 mol/h), i.e. for the "55 bar – 30 °C" system, the TEM micrographs showed scalenohedral calcite as dominant morphology (Figure 4b). For this case, sub-micrometric isolated particles (< 1 µm) and micrometric aggregates and/or agglomerates (< 5 µm) of calcite were observed after 24 h of reaction. A slight proportion of rhombohedral calcite was also observed (Figure 4b). The BET measurements reveal a high specific surface area (9.7 $m^2/g$) compared with typical values of powdered calcite. This value leads to a sub-micrometric average particle size of 0.22 µm assuming spherical or cubic geometry. Obviously, the specific surface area and average particle size of solid matrix depend on the reaction time before equilibrium system (this will be described in the following sub-section). In contrast, at high pressure and temperature, i.e. for the "90 bar – 90 °C" system (precipitation rate = 4.3 mol/h), the TEM micrographs showed rhombohedral calcite as dominant morphology in the product. In this case sub-micrometric isolated particles (< 1 µm) and micrometric agglomerates (< 5 µm) of calcite were observed after 24 h of reaction (Figure 4a). For this case, the specific BET surface area and average particle size were 6 $m^2/g$ and 0.36 µm, respectively.

### 3.3. Rietveld refinements

The results of the Rietveld refinements are shown in Table 1 and the observed and calculated powder diffraction spectra of 2 different samples are presented in Figure 5. Plots of the evolution of the coherent domain average size, maximum and minimum coherent domain size and specific surface area for the calcite samples synthesized under 90 bar of total pressure and 90 °C are presented in Figure 2 for a better interpretation of the results. A residual phase of Portlandite was found is all the refinements up to 4 hours of reaction time; from that reaction time only the calcite phase is present. The growth of the coherent domain size of samples synthesized at $P_{total}$ = 90 bar follows a quick evolution during the first 5 hours of the reaction.

### 3.4. Process of calcite growth

The increase in coherent domain size as detected by X-ray diffraction is accompanied by a precipitation-dissolution process of the crystallites in their very early stage of nucleation (first 0.5 h). This fact can be clearly observed following the kinetics of growing of the crystallites from the XRD coherent domain size shown in Figure 6: it reflects a continuous growing of the minimum size of the coherent domains, while the maximum size decreases from 168 nm to 137 nm when passing from 0.25 h to 0.5 h of reaction. This behaviour is observed later in the reaction: the maximum size of the coherent domains decrease from 412 nm after 15 h of reaction to 337 nm after 24h. This decrease can be interpreted in terms of a dissolution process of the coherent domains, given that there is an excess of $CO_2$ concentration in the reactor leading to an acidic condition which would explain this behaviour. Note that we cannot confirm the same behaviour occurring in the samples synthesized at 50 bar of $P_{CO2}$ and 30 °C as we only extracted samples at 0.25 h and 24 h of reaction.

The experimental BET specific surface area and the calculated average coherent domain surface area resulting from the Rietveld refinements for each sample and experimental synthesis conditions are presented in Table 2. The TEM micrographs in Figure 4 show a well crystallized rhombohedral calcite with BET calculated average sizes in the order of 0.36 µm. The BET calculated average size is obtained by assuming cubic particle shape. The values for the BET and XRD specific surface areas differ by a factor of ~5 for the samples of the "50 bar – 30 °C" synthesis and a factor of ~3 for the "90 bar – 90 °C" synthesis. The values obtained from XRD are always higher than the BET, as expected from the values for the crystallites sizes. The fact that the proportionality factor between BET and XRD goes from 5 to 3 can be interpreted in terms of aggregation state: the coherent domains are less aggregated when the synthesis is done at higher pressure and temperature. TEM observations, even though not giving a statistical mean of the ensemble, reveal crystallite sizes on the same order as the obtained by XRD. These two facts have direct implications on the colloidal behaviour of the calcite as well as in its behaviour with respect to solubility and dissolution processes.

## 4. Conclusion

This study demonstrate that the Pressure, the temperature and the dissolved quantity of $CO_2$ have significant effect on the average particle size, specific surface area, initial rate of precipitation, and on the morphology of calcium carbonate crystals. In contrast, these PTx conditions used herein have insignificant effect on the carbonation efficiency of $Ca(OH)_2$. In addition, molecular $CO_2$ (in gaseous or supercritical state) seems also to have an insignificant effect on the calcite precipitation.

Rietveld analyses of the XRD patterns yield information on the growth behaviour and the crystallinity of the samples. While nanoparticles are obtained at 55 bar and 30 °C, the sub-micrometric well defined crystals are obtained at 90 bar and 90°C. First, a dissolution-precipitation process is observed in the first stages (first hour) of the synthesis. Second, the ratio between the XRD and BET specific surface areas decreases from 5 to 3 when changing the initial thermodynamic conditions from 50 bar and 30 ºC to 90 bar and 90 ºC. This demonstrates that better crystallized samples are obtained when raising the temperature and pressure conditions, and a higher degree of aggregation of the crystallites in the 50 bar and 30 ºC samples. These analyses give useful information on the microstructural characteristics of the calcite samples obtained from hydrothermal conditions.

The results presented here demonstrate that nano-calcite crystals, with high specific surface area can be produced, with a high potential for industrial applications such as adsorbents and/or filler in papermaking industry.

**Acknowledgements**

The authors are grateful to the National Research Agency, ANR (GeoCarbone-CARBONATATION project) and the National Research Council (CNRS), France, for providing a financial support for this work. This study has also been financed through collaboration between the University of Grenoble (German Montes-Hernandez, François Renard) and Gaz de France (Christophe Rigollet, Samuel Saysset, Rémi Dreux). The authors are grateful to S. Petit who allowed the BET measurements to be performed in her laboratory. Nicolas Geoffroy is acknowledged for his help during the XRD experiments.

1   **Table 1.** Results of the Rietveld refinement of powder diffraction data: average crystallite size, strain,

2   maximum and minimum average sizes per crystallite and their reciprocal space directions. The error

3   for all the magnitudes is better than 10%.

| $P_T$ (bar) | Reaction time (h) | XRD coherent domain average size (nm) | XRD coherent domain average strain (%%) | XRD coherent domain maximum size (nm) | [$h\,k\,l$] maximum size | XRD coherent domain minimum size (nm) | [$h\,k\,l$] minimum size | $\chi^2$ |
|---|---|---|---|---|---|---|---|---|
| 55 | 0.25 | 42 | 14 | 54 | [2 0 2] | 35 | [0 1 2] | 10.8 |
|  | 24 | 64 | 14 | 96 | [2 0 2] | 38 | [0 0 6] | 12.7 |
| 90 | 0.25 | 112 | 15 | 168 | [2 0 2] | 75 | [0 1 2] | 8.89 |
|  | 0.5 | 99 | 16 | 137 | [1 0 4] | 76 | [0 1 2] | 9.86 |
|  | 4 | 157 | 17 | 230 | [3 0 0] | 89 | [0 1 8] | 18.2 |
|  | 15 | 157 | 14 | 412 | [0 0 6] | 92 | [0 1 8] | 19.6 |
|  | 24 | 185 | 16 | 337 | [2 0 2] | 109 | [0 1 2] | 14.8 |

7   **Table 2.** Calculated XRD coherent domain specific surface area and experimental BET specific

8   surface area. The BET measurements were performed on 4 out of the 7 available samples.

| $P_T$ (bar) | Reaction time (h) | XRD coherent domain specific surface (m$^2$/g) | BET crystallite specific surface (m$^2$/g) |
|---|---|---|---|
| 55 | 0.25 | 61.3 | 14.55 |
|  | 24 | 42.7 | 9.72 |
| 90 | 0.25 | 24.8 | 7.06 |
|  | 0.5 | 27.9 | - |
|  | 4 | 12.6 | - |
|  | 15 | 14.4 | - |
|  | 24 | 15.3 | 5.95 |

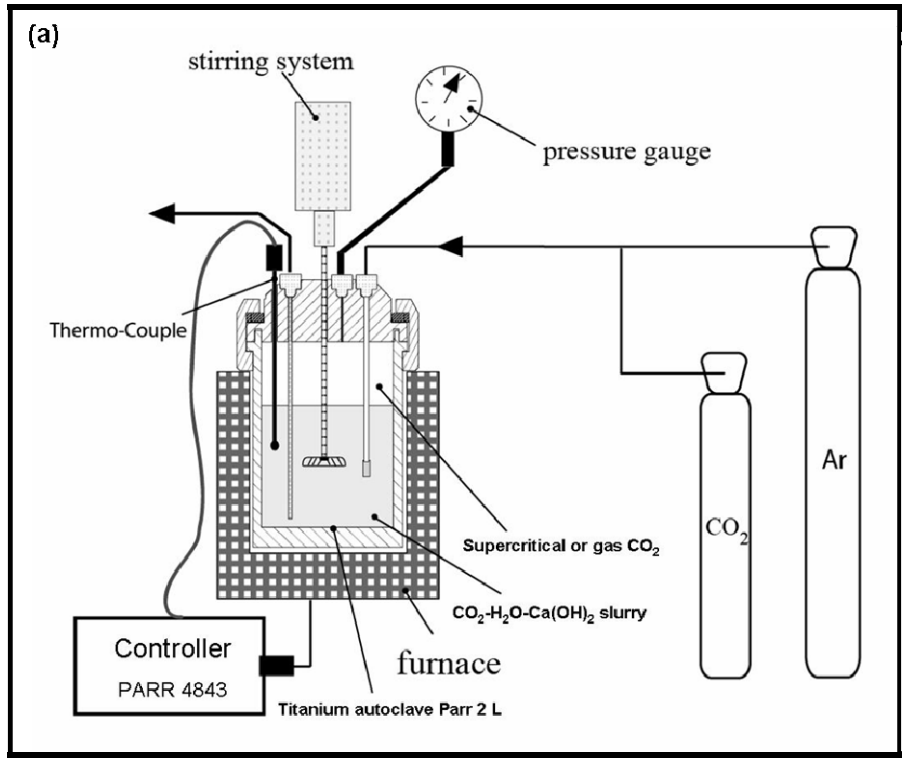

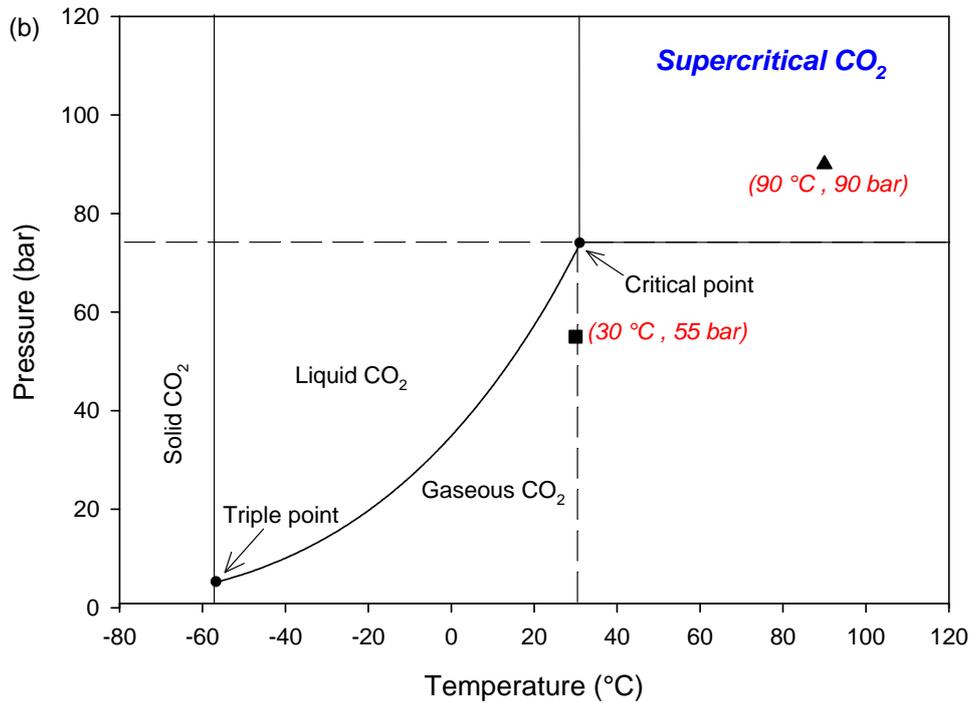

**Figure 1**. (a) Schematic experimental system for calcite precipitation from $CO_2$-$H_2O$-$Ca(OH)_2$ slurry in presence of supercritical and gaseous $CO_2$ (isobaric system). (b) Experimental conditions represented on a pressure-temperature phase diagram for $CO_2$.

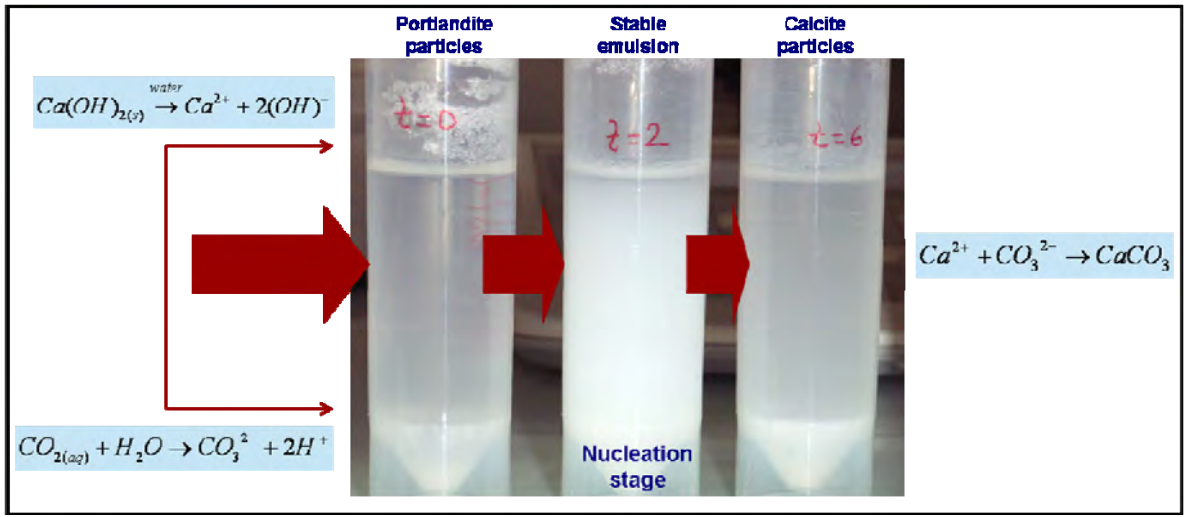

**Figure 2**. Schematic reaction mechanism of calcite precipitation from $CO_2$-$H_2O$-$Ca(OH)_2$ slurry. Suspensions sampled in the reactor before $CO_2$ injection (t=0) and after $CO_2$ injection (t=2 and t=6 minutes).

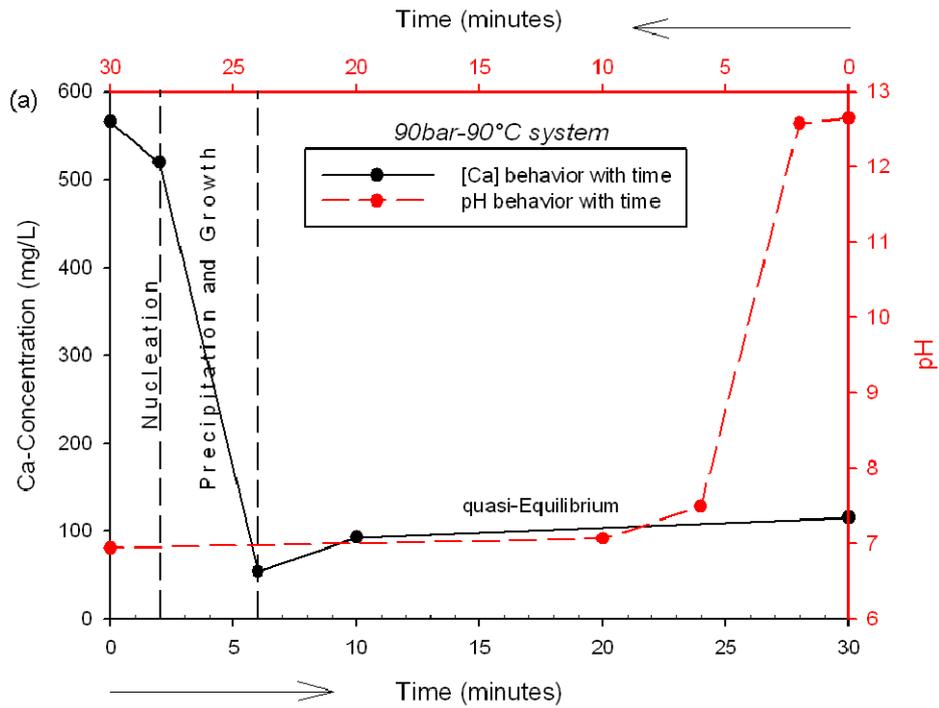

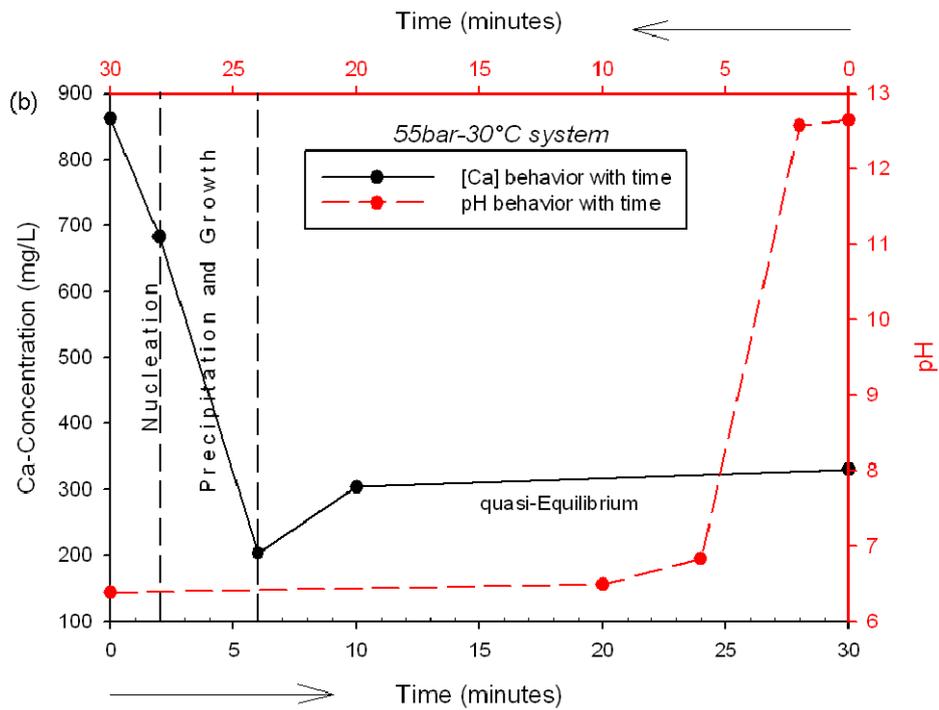

**Figure 3**. Ca-Concentration and pH behaviour with time for calcite precipitation at (a) 90 bar and 90 °C and, (b) 55 bar and 30 °C. Note that the pH and Ca-concentration measurements were carried out at 25 °C after filtration, cooling and degasification of the solutions

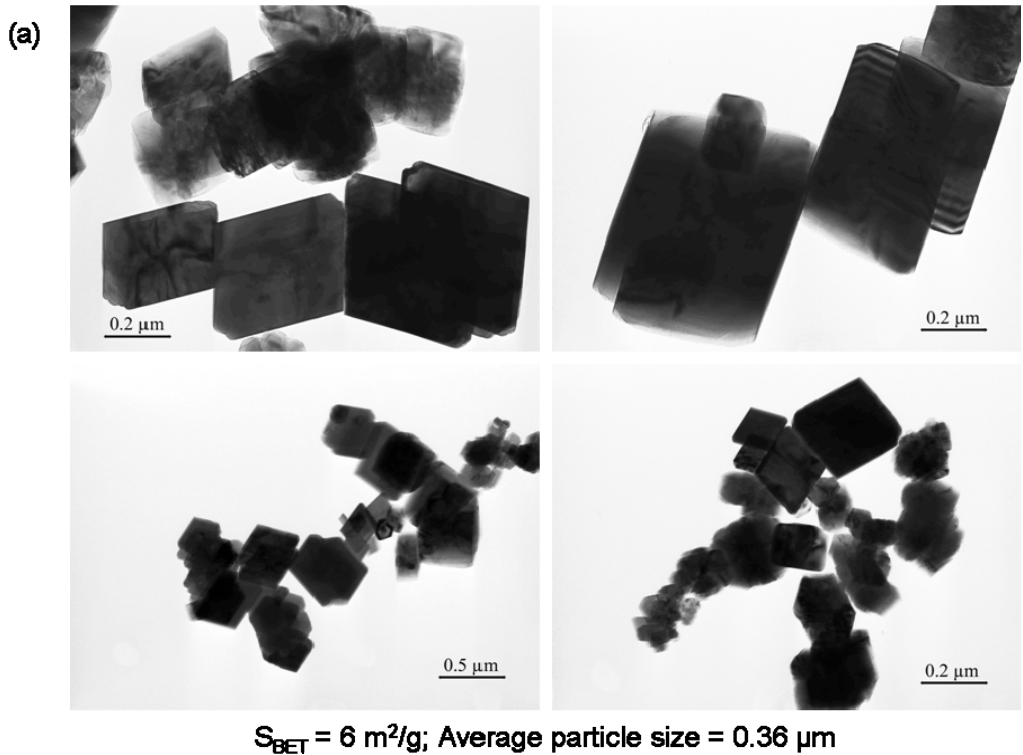

$S_{BET}$ = 6 m²/g; Average particle size = 0.36 μm

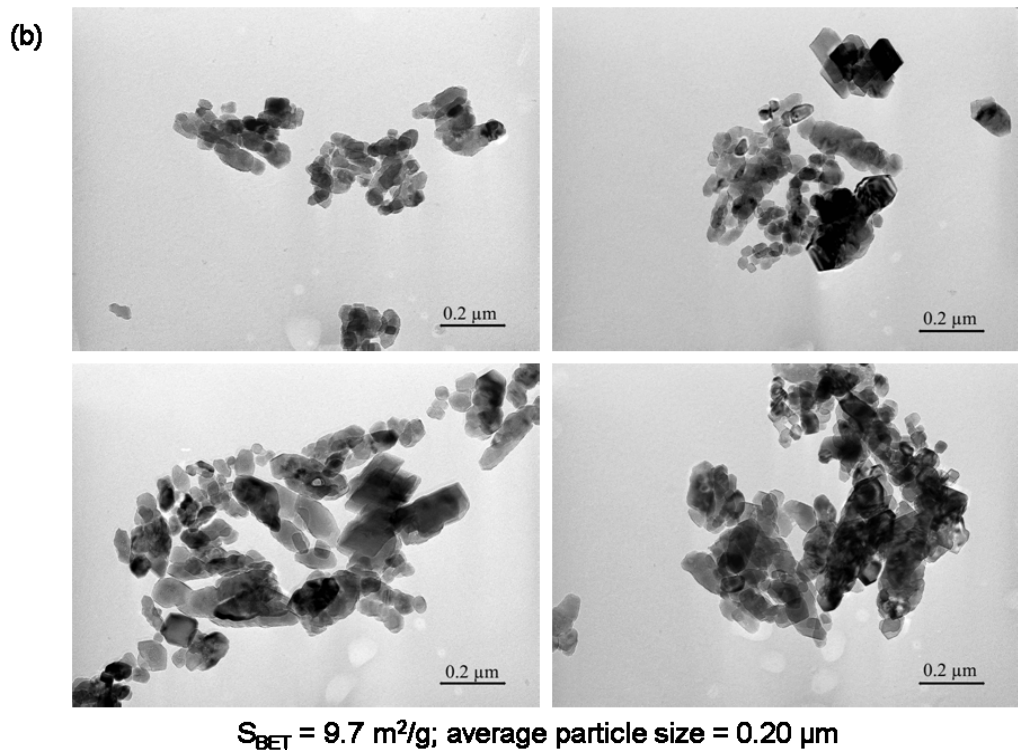

$S_{BET}$ = 9.7 m²/g; average particle size = 0.20 μm

**Figure 4**. TEM micrographs showing the calcite particles precipitated from $CO_2$-$H_2O$-$Ca(OH)_2$ slurry in presence of (a) supercritical $CO_2$ "90 bar; 90 °C" and (b) gaseous $CO_2$ "55 bar, 30 °C" after 24h of reaction in batch system.

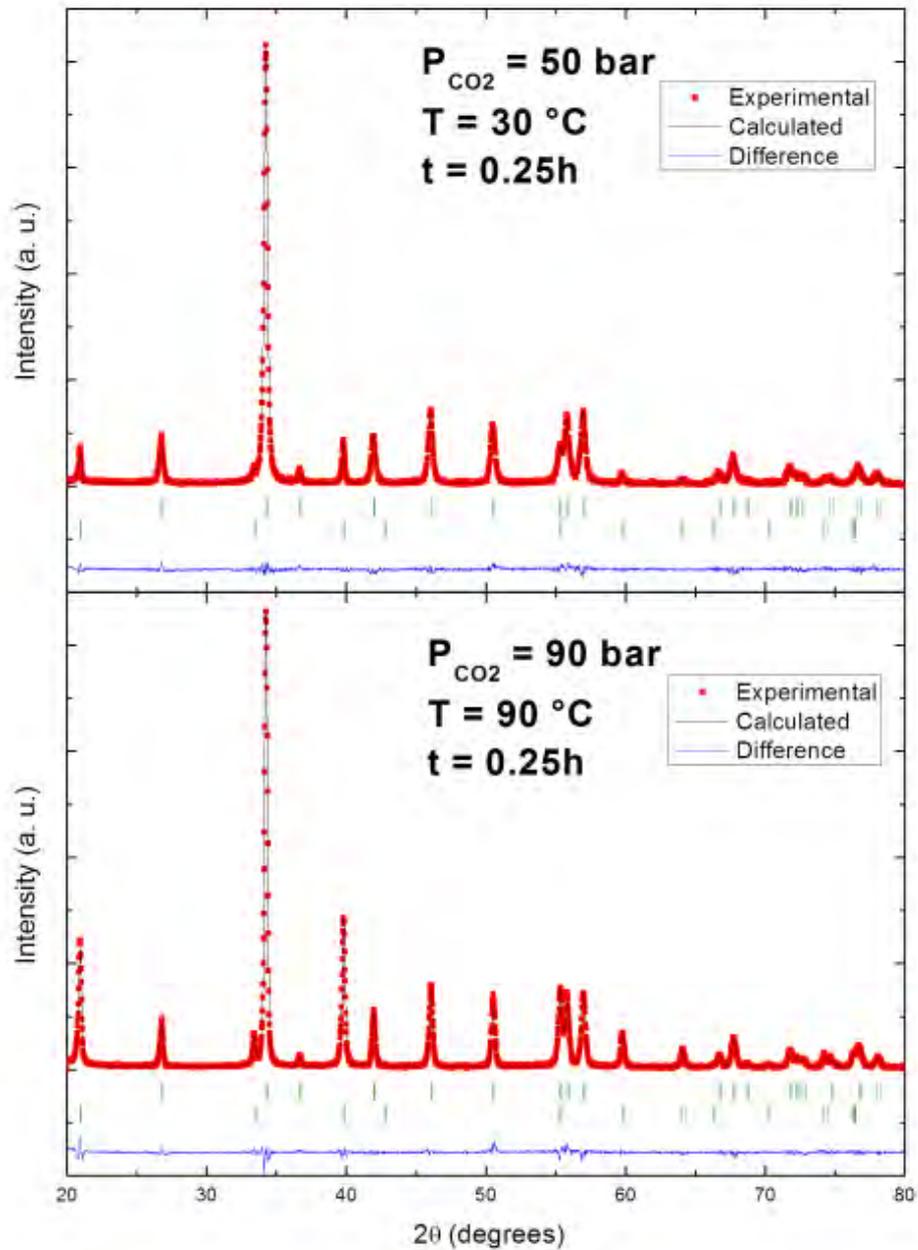

**Figure 5**. Experimental and Rietveld refined X-ray diffraction patterns for the calcite samples synthesized during 0.25 h. The green lines indicate the Bragg reflections. A secondary phase of Portlandite was introduced in all the refinements. An anisotropic spherical harmonics model for the particle size and an isotropic strain model were used to model the width of the Bragg reflections.

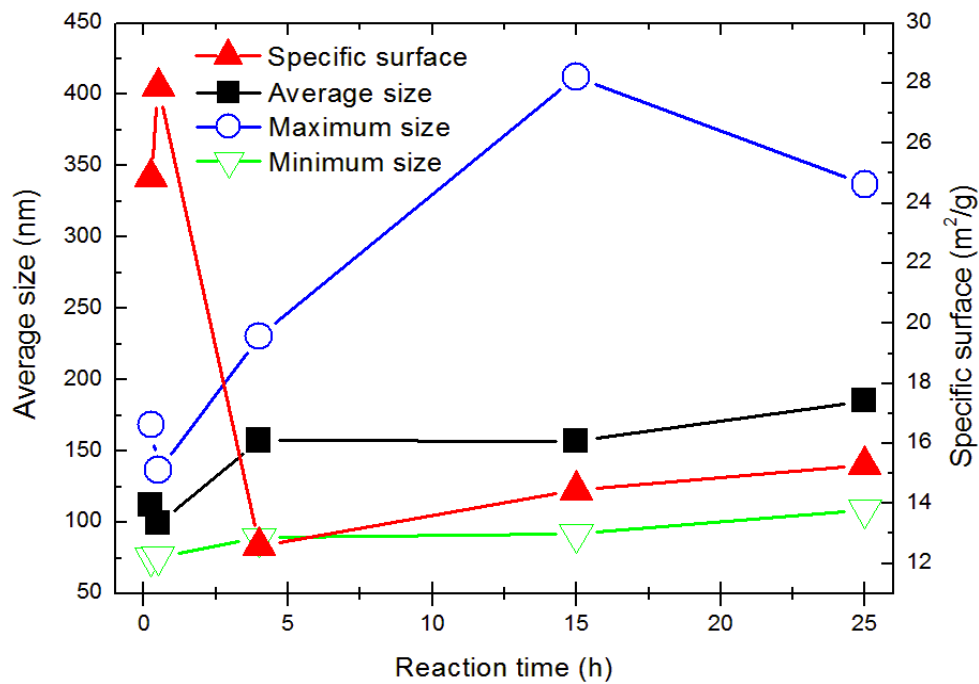

**Figure 6**. Evolution of the coherent domain average, maximum and minimum sizes and of the specific surface area of the calcite samples obtained at 90 bar and 90ºC.